\documentclass[a4paper]{article}
\usepackage[utf8]{inputenc}
\usepackage{pgfplots}
\DeclareUnicodeCharacter{2212}{−}
\usepgfplotslibrary{groupplots,dateplot}
\usetikzlibrary{patterns,shapes.arrows}
\pgfplotsset{compat=newest}
\usepackage{adjustbox}

\usepackage{INTERSPEECH2022}
\usepackage{amsmath}
\DeclareMathOperator*{\argmin}{arg\,min}

\title{ Automatic Data Augmentation Selection and Parametrization in Contrastive Self-Supervised Speech Representation Learning}
\name{Salah Zaiem$^1$, Titouan Parcollet$^2$, Slim Essid$^1$}
\address{
  $^1$LTCI, Télécom Paris, Institut Polytechnique de Paris\\
  $^2$LIA, Avignon Université}
\email{salah.zaiem@telecom-paris.fr}

\begin{document}

\maketitle
\begin{abstract}
Contrastive learning enables learning useful audio and speech representations without ground-truth labels by maximizing the similarity between latent representations of similar signal segments. In this framework various data augmentation techniques are usually exploited to help enforce desired invariances within the learned representations, improving performance on various audio tasks thanks to more robust embeddings. Now, selecting the most relevant augmentations has proven crucial for better downstream performances.  Thus, this work introduces a conditional independance-based method which allows for automatically selecting a suitable distribution on the choice of augmentations and their parametrization from a set of predefined ones, for contrastive self-supervised pre-training. This is performed with respect to a downstream task of interest, hence saving a costly hyper-parameter search. Experiments performed on two different downstream tasks validate the proposed approach showing better results than experimenting without augmentation or with baseline augmentations. We furthermore conduct a qualitative analysis of the automatically selected augmentations and their variation according to the considered final downstream dataset.
\end{abstract}

\section{Introduction and related works}
Self-supervised learning (SSL) enables the use of large amounts of unlabeled data to obtain substantial performance improvements in a wide range of downstream tasks, without relying on costly and maybe imprecise manual annotations. Various approaches have thus been introduced and applied to speech data, including predictive coding \cite{baevski2020wav2vec,Liu_2020}, multi-task learning \cite{ravanelli2020multitask, https://doi.org/10.48550/arxiv.2107.00594}, encoding techniques \cite{algayres:hal-02977539} or contrastive learning \cite{cola,Shor_2020}. \\

\noindent\textbf{Contrastive Learning.}
Specifically, contrastive learning is one of the leading paradigms in speech self-supervised representation learning, especially towards solving paralinguistic classification tasks \cite{shor2021,Al-Tahan2021}. COLA (COntrastive Learning
for Audio) \cite{cola} is an audio-adapted version of these models. It consists in learning representations  through assigning high-similarity to segments extracted from the same audio file and low-similarity to segments from different files. The learned representations are then fed to downstream models solving tasks.
However, unlike similar approaches in the computer vision literature \cite{simclr}, COLA does not explore the use of data augmentation to enforce further invariances in the representations. This work explores this use and its variation with the considered downtream task. \\

\noindent\textbf{Data augmentation in SSL settings.}
In this context, the creation of different versions, often called "views",  of a given data point through data augmentation is an essential part of various self-supervised approaches \cite{simclr, byol}. On speech data, Kharitonov \textit{and al.} \cite{kharitonov} have shown that using data augmentation to alter the data during Contrastive Predictive Coding (CPC) \cite{rivire2020unsupervised, Oord2018RepresentationLW} training improves the downstream ASR performance. Two works may be considered as close to the purpose of this paper. First, in image classification settings, adapting the augmentation distribution used in the contrastive pretraining to the downstream classification task has proven effective \cite{Xiao2020, li2021selfsupervised}. This is particularly true when certain differences, to which the representations are trained to be invariant, are crucial for distinguishing the downstream classes. Second, experiments led on contrastive representations (COLA based) on sound classification show that augmenting the cut segments leads to better results, and that the set of best performing augmentations is downstream task dependent \cite{Emami2021}.  Nonetheless, while ablation studies are conducted on the selected augmentations, no prior justification of the choices are developed, making the selection relying on computationally heavy empirical exploration.
\begin{figure*}[ht!]
  \centering
  \includegraphics[width=0.8\linewidth, scale=0.2]{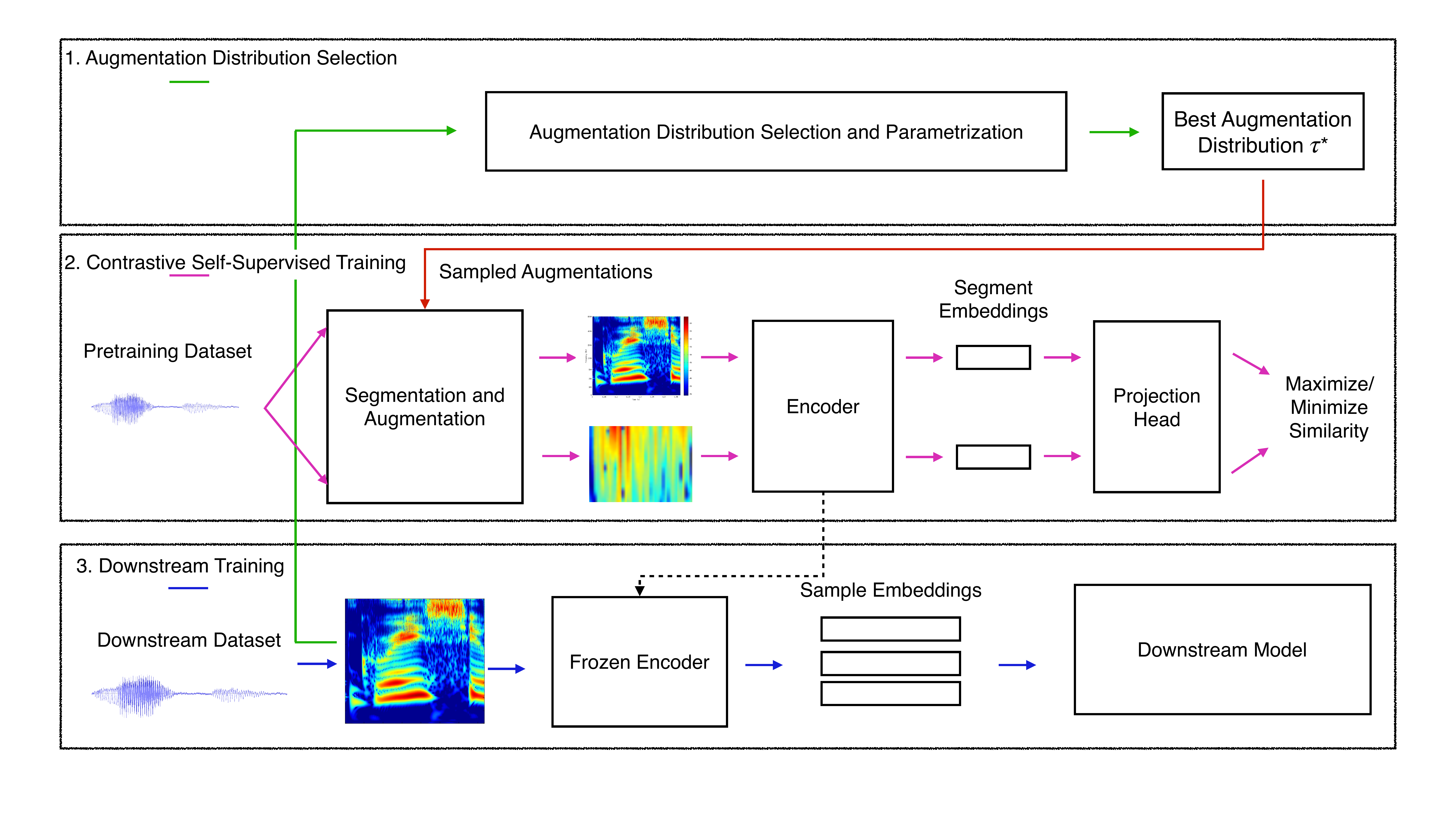}
  \vspace{-0.5cm}
  \caption{The three steps of the validation process. (a) select the best augmentation distribution. (b) contrastive pretraining alterating the input points with the selected augmentation. (c) use the learned speech representations as input for downstream finetuning.}
  \vspace{-0.5cm}
  \label{fig:schema}
\end{figure*}
Finally, a few works have attempted  to define how views should be created in contrastive learning settings \cite{Arora2019, tian2020makes},  and thus which and how augmentations should be used. However, and to the best of our knowledge, there is no attempt to theoretically motivate data augmentation in self-supervised settings on speech or audio data.  This work will rely on COLA approach as it is one of the closest to vanilla contrastive learning, and it did not explore the use of data augmentation on speech. It is, nonetheless, perfectly transferable to other contrastive approaches.  If we were to rely only on empirical testing, evaluating a single set of augmentation distribution would require two full trainings. In the specific case of this paper, a single pretraining takes 2 days on a V100 GPU. The method we present prevents this, allowing for an efficient selection of an appropriate data augmentation distribution.  
The contributions of this work are thus threefold : 
\begin{enumerate}
    \item To highlight the impact of data augmentation on contrastive self-supervised speech representation learning.
    \item To propose a method that selects a distribution on the choice of augmentations and their parametrization according to the downstream task of interest, validated on two different downstream tasks. The selected augmentations are qualitatively linked to the datasets recording conditions.
    \item To release the code base, implemented with SpeechBrain \cite{speechbrain} for replication and further improvements. \footnote{\url{https://github.com/salah-zaiem/augmentations}}
\end{enumerate}

Figure \ref{fig:schema} presents an overview of the led experiments, summarizing the three steps conducted for every downstream task. First, an augmentation distribution  is selected (Section \ref{sec:selection}). Second, representations are learned through contrastive pre-training using the selected augmentation distribution (Section \ref{sec:contrastivelearning}). Finally, the learned representations are fed to the downstream model to solve the considered task (Section \ref{sec:downstream}).

\section{Selecting the Augmentation Distribution}
\label{sec:selection}
This section details the method developed to find a  data augmentation distribution for the contrastive learning part, suitable to the final downstream task of interest. It starts by detailing the theoretical motivations behind the method, before delving into the technical details of the implementation. 

\subsection{Theoretical Motivation}
During self-supervised training, the representations are learned through solving automatically generated pretext tasks. Lee \textit{and al.} \cite{lee2020predicting}  theoretically proved a link between the downstream task performance and the conditional independence (CI) between the pretext task labels and the training samples given the downstream labels. In previous works \cite{Zaiem_2021,https://doi.org/10.48550/arxiv.2107.00594}, we have shown that this relation holds even without the theoretical assumptions in \cite{lee2020predicting}, introducing a practical method to compute the conditional independence. Precisely, let $X$, $Y$ and $Z$ be, respectively, the downstream data points, the downstream labels and the pretext labels whose prediction is used as a pretext task. We have shown that the more is $Z$ conditionally independent of $X$ given $Y$, the more using the prediction of $Z$ as a pretext task leads to a better downstream performance. To quantify the utility of a given pretext task, we use the kernelized independence test Hilbert Schmidt Independence Criterion (HSIC) \cite{NIPS2007_d5cfead9}. Intuitively, the HSIC value is high if similar speech samples have similar pretext labels. The more the HSIC value is high, the more dependent are $X$ and $Z$ conditionally on $Y$. In \cite{Zaiem_2021,https://doi.org/10.48550/arxiv.2107.00594}, we demonstrated that choosing the pretext labels $Z$  minimizing $HSIC(Z,X | Y)$ leads to better downstream performances. 

In this work, we extend these findings to the contrastive learning settings through the following steps. First, the key consists in considering that in the contrastive learning setting, the pretext task of assigning high similarity to segments originating from the same file can be seen as the prediction, given a random augmented segment, of the file it was generated from. An augmentation distribution $\tau$ is defined by a set of parameters defining how a chain of augmentations is sampled during training to be applied to the upcoming data points. More precisely, every distribution $\tau$ is represented as a vector of $P=14$ parameters, where every parameter $(\tau(p))_{1\leq p \leq P}$ is either the probability of applying an augmentation or a boundary for a uniform law from which a augmentation's internal parameter (e.g. room scale) is sampled. With $X$ the speech samples and $\tau$ a distribution of augmentations, we define $X' = f(X, \tau)$ with $f$ a function that randomly cuts segments from the speech samples and applies augmentations sampled from $\tau$ on them.  Given a downstream dataset of samples $(X,Y)$ and an augmentation distribution $\tau$, we can generate $N$ augmented segments per speech sample to get the augmented set of data points $X'$. To find the optimal augmentation distribution $\tau^*$ we resort to minimizing the HSIC quantity with the augmented dataset $X' = f(X, \tau)$ according to:
\begin{equation}
\label{minimize}
    \tau^* = \argmin_\tau HSIC (f(X,\tau), Z | Y)
\end{equation}
with $(X,Y)$ the downstream datapoints and labels, and $Z$ the pretext labels corresponding here for every augmented view of a speech sample to the ID of the speech sample it originates from.

\subsection{Implementation}
In this work, we chose to limit ourselves to the set of augmentations used in \cite{kharitonov} for two reasons. First, they have shown effective with the contrastive predictive coding approach improving the final discrimination performances. Second, they are easily implemented within PyTorch using the WavAugment library. Hence, five augmentations are considered: time dropping\cite{Park_2019}, pitch shifting \cite{10.2307/3679554}, reverberation, clipping and band rejection\cite{Park_2019}. The first parameters concern the probability of applying each one of the considered augmentations. The second set of parameters are related to those of the chosen augmentations in terms of signal effects; these are described in Table \ref{tab:parameters}. 

Since the considered augmentations are not differentiable, to minimize the HSIC test described above, we resort to a random search, sampling random distributions and selecting the one with the lowest dependance scoring. It is important to note here, that this phase does not involve any training, and is largely more efficient than thorough testing of the distributions, as a computation takes 3 hours on 20 CPUs. More precisely, for every considered downstream task, we first sample $p=100$ parametrizations $(\tau_i)_{i \in [1,p]}$. For every parametrization $\tau_i$, we compute the HSIC quantity in Eq.\eqref{minimize} following two steps. First, computing the augmented set $X'_i = f(X, \tau_i)$, by computing $N=20$ views of every speech sample in $X$. Then, computing $HSIC(X', Z | Y)$ following the technique described in \cite{Zaiem_2021}. For every downstream task, the augmentation distribution with the lowest conditional dependance value is selected and will be used during the pretraining to train the encoder that will be exploited as a feature extractor in the downstream training.

\begin{table}[th]
  \caption{Parameters considered, descriptions and ranges}
  \label{tab:parameters}
  \centering
  \scalebox{0.8}{\begin{tabular}{lll}
    \toprule
    \multicolumn{1}{l}{\textbf{Name}} & 
                                         \multicolumn{1}{l}{\textbf{Description}}
                                     &    \multicolumn{1}{l}{\textbf{Range}}\\
    \midrule
                              Room scale min& Min room size              &  [0,30]\\
     Room scale max & Max room size  & [30,100]              \\

                         Band Scaler    & Scales the rejected band & [0,1]            \\
     Pitch Shift Max &Amplitude of a pitch shift & [150,450]             \\
     Pitch Quick pr. & Speeds pitch shifting&[0,1] \\
     Clip Min&Minimal clip factor & [0.3, 0.6]              \\
     Clip Max&Maximal clip factor  & [0.6, 1]         \\
     Timedrop max & Size of a time dropout & [30-150] ms\\

    \bottomrule
  \end{tabular}}
  
\end{table}
\section{Experimental setup}
This section describes the experiments led to validate the proposed approach and the selected augmentation distributions. It starts by describing the details of the contrastive learning phase before reporting the downstream finetuning conditions.

\subsection{Contrastive Learning}
\label{sec:contrastivelearning}
As shown in Figure \ref{fig:schema}, during the contrastive pre-training, we start by extracting two random segments from every speech sample of a given batch. These segments are then alterated using the considered augmentation distribution before being fed to the encoder. 
Our pretraining model takes as input the speech samples as $64$-Mel band spectrograms. The frame size is $25$ms and hop size $10$ms. As in COLA, the encoder is an  EfficientNet-B0 \cite{efficientnet}, a lightweight convolutional neural network. We cut from the input speech samples  1-second long segments that are augmented using the considered augmentation distribution. Fixing the length of the extracted segments allows the use of EfficientNet-B0 even though it has been originally proposed for computer vision, as fixed length Mel-spectrograms have a 2D structure similar to image inputs. The encoder applies a global time-pooling at its final layer to get a $1280$-dimensioned embedding $h$ that represents the whole segment and that will be the one used for downstream finetuning. During the pretraining phase, this embedding is then projected with a dense layer followed by a layer normalization and a hyperbolic tangent activation to a $512$-sized vector $v$. Learning consists in maximizing the similarity of segments originating from the same file, while minimizing that of those that do not. As suggested by the final results obtained with COLA, the similarity is computed using the bilinear similarity. More precisely, if $g$ is the function regrouping the encoder and the projection head, $x_1$ and $x_2$ two speech segments and $W$ the bilinear parameters, then the similarity function is  $
    s(x_1, x_2) = g(x_1)^T \: W \: g(x_2)$. 
The input is a batch of size $B$ of distinct speech files that we denote $(x_i) _{i \in [1,B]}$, and a selected augmentation distribution $\tau$ from which we can sample at each iteration two augmentation functions $A_\tau$ and $A'_\tau$. From each speech sample, two random segments of length 1 second are cut. The first is altered using  $A_\tau$ while the second undergoes  the $A'_\tau$ alteration, leading to two sets $(\tilde{x}_i)_{i \in [1,B]}$ and $(\tilde{x}'_i)_{i \in [1,B]}$. Finally, the loss function for pretraining is the multi-class cross entropy over the bilinear similarity scores: 
\begin{equation}
    \mathcal{L} = - \log \frac{e^{s(\tilde{x}_i, \tilde{x}'_i)}}{e^{s(\tilde{x}_i, \tilde{x}'_i)} + \sum_{j \ne i}  e^{s(\tilde{x}_i, \tilde{x}'_j)}}.
\end{equation}

\noindent\textbf{Pretraining dataset.} The train set of the English Common Voice dataset (version $8.0$) \cite{ardila2020common} is used for SSL pretraining ($2185$ hours). Common Voice is a collection of speech utterances from worldwide users recording themselves from their own devices. Hence, the closeness to natural settings makes it a suitable choice for self-supervised learning. We remove from Common Voice the sentences lasting more than $10$ seconds, as they often contain long silence parts due to open microphones. It is important to note that since the COLA embeddings were originally introduced to set non-speech tasks as well, they were trained on AudioSet \cite{audioset}, which contains speech and non-speech utterances. Since we will be only working on speech downstream tasks, we chose to use a speech-only pretraining. We also  use a 1024 batch size. All the models are pre-trained for 100 epochs with ADAM and a $10^{-4}$ learning rate. 

\subsection{Downstream finetuning}
\label{sec:downstream}
Two downstream tasks are considered in this work: speaker identification and language identification. Two reasons motivate this choice. First, among the list of tasks COLA was applied on, we chose the two downstream tasks exhibiting the largest room for improvement. Second, we wanted two tasks that would require different aspects of the considered speech signal, thus maybe requiring different sets of augmentations. A study validating this assumption is provided in Section \ref{sec:discussion}. 
VoxCeleb1 \cite{Nagrani_2017} is used for the speaker recognition task. The training set contains $148,642$ utterances from $1251$ different speakers. We use the available identification split for testing. 
VoxForge \cite{maclean2018voxforge} is used for language identification. 6 European languages are present in the 176,438 samples of the dataset, two tenths are kept for validation and testing. 

\begin {figure*}[!hbtp]
\centering
\caption{Difference of the probability of picking an augmentation between the best and worst scoring augmentations, depending on the downstream dataset. Green bars show augmentations that are more likely to get picked for the best scoring distributions for that task. For instance, the far right bars indicate that clipping is an encouraged augmentation on VoxForge, and is discouraged on VoxCeleb1.}

\begin{adjustbox}{scale=0.55}
\begin{tikzpicture}

\definecolor{darkgray176}{RGB}{176,176,176}
\definecolor{green01270}{RGB}{0,127,0}

\begin{groupplot}[group style={group size=2 by 1}] 
\nextgroupplot[
tick align=outside,
tick pos=left,
title={Language ID - Individual recordings},
x grid style={darkgray176},
xmin=-0.475, xmax=4.475,
xtick style={color=black},
xtick={0,1,2,3,4},
xticklabels={bandreject,pitch,reverb,time\_drop,clip},
y grid style={darkgray176},
ymin=-0.208527531074899, ymax=0.225862548157444,
ytick style={color=black}
]
\draw[draw=none,fill=green01270] (axis cs:0.75,0) rectangle (axis cs:1.25,0.184049957109641);
\draw[draw=none,fill=green01270] (axis cs:1.75,0) rectangle (axis cs:2.25,0.0902786280197447);
\draw[draw=none,fill=green01270] (axis cs:3.75,0) rectangle (axis cs:4.25,0.132989939735951);
\draw[draw=none,fill=red] (axis cs:-0.25,0) rectangle (axis cs:0.25,-0.0612691028187037);
\draw[draw=none,fill=red] (axis cs:2.75,0) rectangle (axis cs:3.25,-0.019693863922605);

\nextgroupplot[
scaled y ticks=manual:{}{\pgfmathparse{#1}},
tick align=outside,
tick pos=left,
title={Speaker ID - Professional recordings},
x grid style={darkgray176},
xmin=-0.475, xmax=4.475,
xtick style={color=black},
xtick={0,1,2,3,4},
xticklabels={bandreject,pitch,reverb,time\_drop,clip},
y grid style={darkgray176},
 ylabel={MED},
ymin=-0.208527531074899, ymax=0.225862548157444,
ytick style={color=black},
yticklabels={}
]
\draw[draw=none,fill=green01270] (axis cs:0.75,0) rectangle (axis cs:1.25,0.206117544555974);
\draw[draw=none,fill=green01270] (axis cs:2.75,0) rectangle (axis cs:3.25,0.0632253552074544);
\draw[draw=none,fill=red] (axis cs:-0.25,0) rectangle (axis cs:0.25,-0.0429262939609826);
\draw[draw=none,fill=red] (axis cs:1.75,0) rectangle (axis cs:2.25,-0.0729764421020777);
\draw[draw=none,fill=red] (axis cs:3.75,0) rectangle (axis cs:4.25,-0.188782527473428);
\end{groupplot}

\end{tikzpicture}
\label{fig:probs}

\end{adjustbox}
\end{figure*}
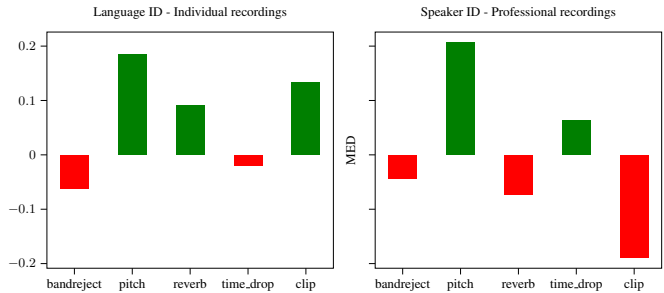

During the downstream finetuning the projection head is discarded and replaced with a linear classifier directly on top of the encoder. The contrastive encoder is frozen during the finetuning phase as we want it to be used solely as a fixed feature extractor to properly assess the impact of our data augmentation selection on the obtained representation. In the COLA paper, the final class prediction is obtained through averaging the predictions of non-overlapping cut segments of a given test utterance. However, we found it more effective to use the mean over the embeddings of overlapping segments. We proceed in this manner: during training and testing, we cut a $1$-s long segment every $200$ms , encode every segment separately, and then use the mean over the encoded representations as a sequence embedding to the classifier. We train on the downstream task for $10$ epochs with ADAM with a $10^{-3}$ learning rate and the additive angular margin loss \cite{arcface} with margin $0.2$ and scale $30$. 

\section{Results and Discussion}
Table \ref{tab:results} shows the results obtained on the two considered downstream tasks. The ``COLA'' column shows the results obtained in the original paper. The ``Without'' column is our implementation of the algorithm without any augmentation during pretraining. ``Basic'' shows the results reached using the baseline WavAugment augmentation parameters. Finally, the results obtained using the augmentation choice based on the proposed technique can be found in the ``Selected'' column. The first observation is that the selected augmentation technique outperforms the baselines on the two considered tasks. For speaker identification, the accuracy obtained with the selected distribution is $46$\% higher than the non-augmented COLA, and $4$\% higher than the baseline augmentations. An important point is that in the baseline augmentation setup (i.e. ``Basic''), all the augmentations are systematically performed on the input points, thus considerably slowing the pretraining. Indeed, with WavAugment augmentations being CPU-processed, we witnessed that dividing by half the conducted augmentations by lowering their probability, leads to 20\% faster trainings. 

\subsection{Discussion}
\label{sec:discussion}
In this part, we will discuss the automatically selected data augmentations, and analyze their dependence on the downstream dataset. We will study first the dependence of the probabilities of applying a given augmentation according to the downstream dataset of interest. Then, we will consider the choice of a few interpretable parameters. This is done through the following procedure: for every downstream task, we start by selecting $k=10$ best and worst augmentation distributions according to our HSIC scoring. The ``Mean Extremal Difference'' or ``MED'' is finally obtained by computing the difference between the two means originating from these two groups \textit{i.e.}, best and worst. More precisely, for an augmentation parameter $p$: 
\begin{equation}
    MED(p)=  \frac{1}{k} ( \sum_{i=0}^k \tau_i^{best}(p) - \tau_i^{worst}(p))
\end{equation} with $\tau_i^{best}$ being the $i$-th best distribution, $\tau_i^{worst}$ being the $i$-th worst and $\tau(p)$ being the value of parameter $p$ in $\tau$.
\begin{figure}[t]
  \centering
  \includegraphics[width=\linewidth, height=3.5cm, keepaspectratio]{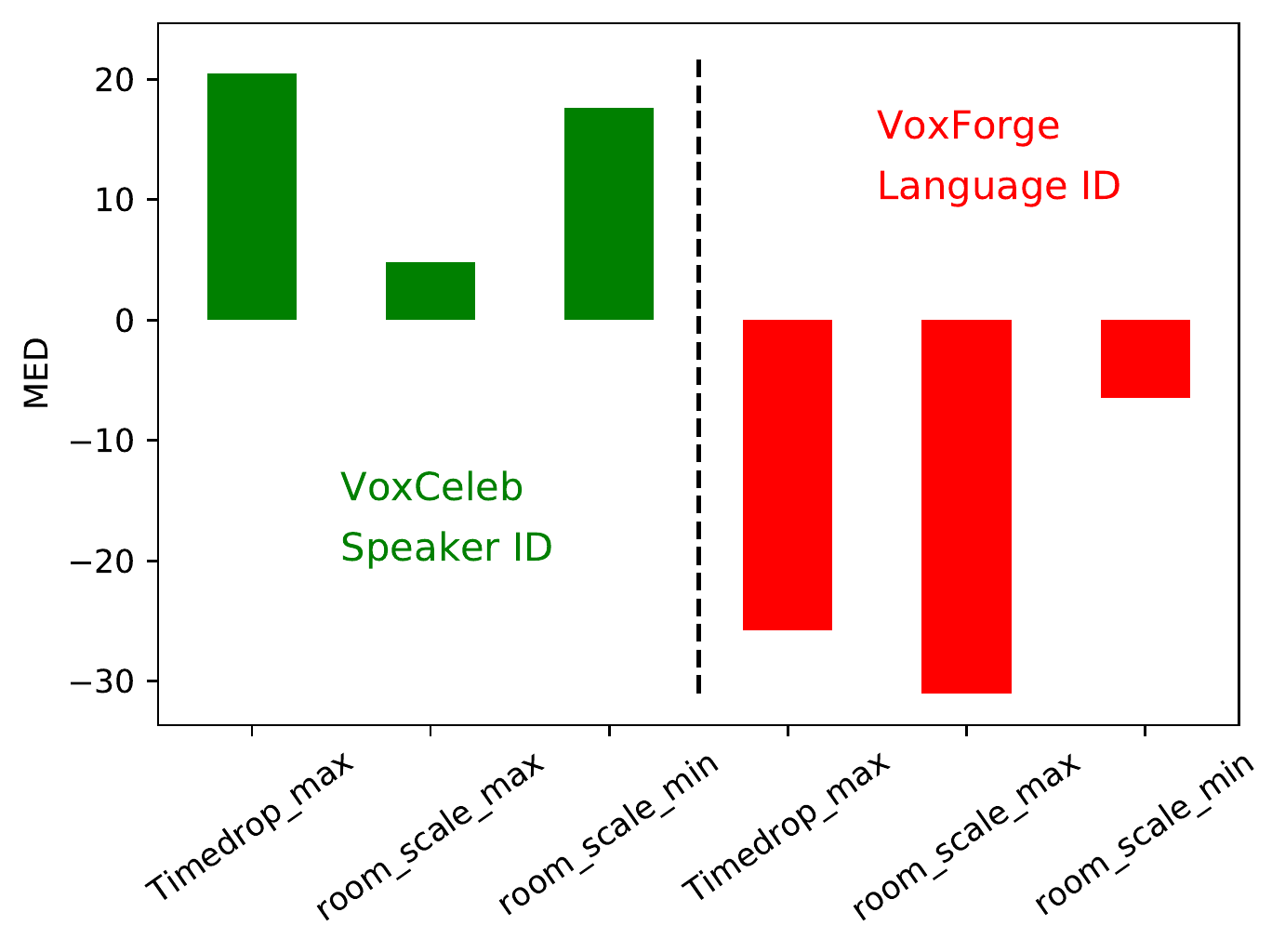}
  \caption{MED for selected parameters, for every downstream dataset. Reverb room sizes are coherent with the difference in recording conditions between the two datasets.}
  \label{fig:values}
\end{figure}

Figure \ref{fig:probs} depicts these values for the probabilities of applying each of the five considered alterations in this work. Green bar means are for positive values, indicating that this augmentation is more likely to be applied in the supposedly best distributions. We observe that clipping and reverberation are more selected for language identification on VoxForge than for speaker identification on VoxCeleb. We think that this is mainly due to the type of recording rather than to the nature of the task. VoxForge samples come from individual contributors that record themselves speaking their native language. The varying recording conditions lead to clipping or heavy reverberation issues, which may be the reason behind the selection of these augmentations in this case. Figure \ref{fig:values} shows the mean difference defined above on $3$ parameters, which are time dropping and room scale boundaries. Concerning reverberation, it is worth noting that room scales are smaller for VoxForge than for VoxCeleb1, which is once again coherent with the recording conditions, as the first ones are recorded at home, compared to studio conditions. Samples of augmented speech files with various distributions are provided for quantitative comparison by the readers.\footnote{\url{https://salah-zaiem.github.io/augmentedsamples/}}

\begin{table}[t!]
  \caption{Results for the two considered downstream tasks. COLA column shows the result of the original paper. "Basic" shows the result with the basic WavAugment recipe. "Selected" shows our approach results.}
  \label{tab:results}
  \centering
  \scalebox{0.8}{\begin{tabular}{*{5}{c}}
    \toprule
    \multicolumn{1}{c}{\textbf{Down. Task}} &  \multicolumn{1}{c}{\textbf{COLA}} 
                                          &\multicolumn{3}{c}{\textbf{Our Implementations}}
                                         \\
    &&Without & Basic & Selected \\
    
    \hline
    \rule{0pt}{2.5ex}
Language ID & 71.3& 84.9 & 84.3& \textbf{85.2} \\
Speaker ID & 29.9 & 32.0& 45.1& \textbf{46.9} \\ 
    \bottomrule
  \end{tabular}}
\end{table}
\section{Conclusion}
Self-supervised learning of speech representations is a computationally intensive technology, especially when using data augmentation within contrastive schemes. To avoid wasting large amounts of resources and to further improve the results observed with such models on various downstream tasks of interest, we introduced a novel informed  method enabling the automatic selection and parametrization of the crucial data augmentation pipeline. Our findings open a range of possibilities in signal alterations exploration for self-supervision.

\bibliographystyle{IEEEtran}
\bibliography{mybib}
\end{document}